\documentclass[11pt]{cernrep}
\usepackage{graphicx}
\usepackage{here}

\def \ket#1{| #1 \rangle}

\begin{document}

\title{
\vskip -25mm
\hfill \small DESY 02-194 \\
\vskip 15mm
THE PHENOMENOLOGY OF GLUEBALL AND HYBRID MESONS}
\author{Stephen Godfrey}
\institute{Department of Physics,
Carleton University, Ottawa K1S 5B6 CANADA \\
DESY, Deutsches Elektronen-Synchrotron, D22603 Hamburg, GERMANY}
\maketitle
\begin{abstract}
The existence of non-$q\bar{q}$ hadrons such as glueballs and hybrids 
is one of the most important qualitative questions in QCD.  The 
COMPASS experiment offers the possibility to unambiguously identify
such states and map out the glueball and hybrid spectrum.  In this 
review I discuss the expected properties of glueballs and hybrids and 
how they might be produced and studied by the COMPASS collaboration.
\end{abstract}

\section{INTRODUCTION}

A fundamental qualitative question in the Standard Model is the 
understanding of quark and gluon confinement in Quantum 
Chromodynamics.  Meson spectroscopy offers the ideal laboratory to 
understand this question 
which is intimately related to the question of ``How does glue 
manifest itself in the soft QCD regime?'' \footnote{For a more 
detailed review on this subject see Ref. \cite{gn99}. }
 Models of hadron structure 
predict new forms of hadronic matter with explicit glue degrees of 
freedom:  Glueballs and Hybrids.  The former is a type of 
hadron with no valence quark content, only glue, while the latter has 
quarks and antiquarks with an excited gluonic degree of freedom.  In 
addition, multiquark states are also 
expected.  With all these ingredients, the physical spectrum is 
expected to be very complicated.

Over the last decade there has been considerable theoretical progress 
in calculating hadron properties from first principles using Lattice 
QCD \cite{lattice,michael00}.  
This approach gives a good description of the observed spectrum 
of heavy quarkonium and supports the potential model description, at least 
for the case of heavy quarkonium.  

Lattice QCD now has reasonably robust predictions for glueball masses
\cite{lattice,gbmass,mp97}, 
albeat in the quenched approximation.  Although there is growing 
evidence for the observation of glueballs it has required considerable 
theoretical analysis to argue that there is an extra isoscalar 
$J^{PC}=0^{++}$ state in the meson spectrum.  The problem is that a 
glueball with quantum numbers consistent with those of conventional 
$q\bar{q}$ mesons will mix with the $q\bar{q}$ states
complicating the analysis of their couplings \cite{ca96}.  
There is a strong need 
to unambiguously observe glueballs and perform detailed analysis of 
their properties as a rigorous test of QCD.  A deeper reason 
for these studies is that lattice field theory has become an important tool 
for understanding strongly coupled field theories.  QCD is 
the one place where we can test our calculations against experiment so 
that agreement with measurements will give us the confidence that we 
really can do nonperturbative field theory calculations.

Hybrid mesons pose another important test of our understanding of 
QCD.  It is now clear that lattice QCD calculations support the flux 
tube picture of hadron dynamics, at least in the heavy quark limit 
\cite{bali94}.  
Excitations of the flux tube are described by non-trivial 
representations of the flux tube symmetry \cite{jkm98}.  
A good analogy is 
that of the electron wavefunctions in diatomic molecules.  In this 
picture,
conventional mesons are described by a $q\bar{q}$ potential given by
the lowest adiabatic surface and hybrids are described by a $q\bar{q}$ 
potential given by higher adiabatic surfaces arising from different 
flux tube symmetries.  It is necessary to map out these higher adiabatic 
surfaces to test our understanding of ``soft QCD''.  To do so requires 
the observation of enough states to map out these excited surfaces.

Although lattice calculations are maturing, giving more reliable 
results for masses, it will be some time before they can reliably 
describe decay and production couplings.  We therefore rely on
phenomenological models to describe their properties and build up a 
physical picture needed to help find these states.

\section{CONVENTIONAL MESONS}

To search for glueballs and hybrids it is necessary to have reliable 
descriptions of conventional mesons \cite{gi,bcps,bbp}.  
Conventional mesons are composed 
of a quark-antiquark pair.  The various quark flavours are combined 
with antiquarks to form the different mesons.  The meson quantum numbers 
are characterized by a given $J^{PC}$.  In the constituent quark model
the quark and antiquark spins 
are combined to give a total spin with $S=$0, 1.  $S$ is then combined 
with the orbital angular momentum $L$ to give total angular momentum 
$J=L+S$.  Parity is given by $(-1)^{L+1}$ and charge conjugation by 
$C=(-1)^{L+S}$.  This results in allowed quantum numbers, for example, 
$J^{PC}=0^{-+}$, $1^{--}$, $1^{+-}$, $0^{++}$, $2^{++} \ldots$ while  
$J^{PC}=0^{--}$, $0^{+-}$, $1^{-+}$, $2^{+-}$ are forbidden by the 
quark model and are generally referred to as exotics.  

Although the goal is to discover non $q\bar{q}$ states we can't ignore 
conventional mesons.  We need to understand them quite well if we are 
to disentangle the non-$q\bar{q}$ states we seek from conventional 
$q\bar{q}$ mesons .  We 
can do this because the couplings of states are sensitive to their 
internal structure.  Strong decays are modeled by the $^3P_0$ model 
and by the flux-tube breaking model \cite{bcps,bbp} 
while  electromagnetic couplings 
are quite well understood for heavy quarkonium and qualitatively for 
light quark mesons.  The electromagnetic couplings can be measured in 
$2\gamma$ couplings and single photon transitions.  The latter can be
measured via Primakoff production by COMPASS.

\section{GLUEBALLS}

The predictions of glueball masses by Lattice QCD are becoming fairly 
robust\cite{michael00}. The results of a Lattice QCD calculation of the glueball 
spectrum by Morningstar and Peardon \cite{mp97} are given in Fig. 1.  
The lowest 
mass glueballs have conventional quantum numbers \cite{gbmass}:
$M_{0^{++}}\sim 1.6$~GeV, $M_{2^{++}}\sim 2.3$~GeV 
$M_{0^{-+}}\sim 2.5$~GeV while the lowest lying glueballs with exotic 
quantum numbers,
$J^{PC}=0^{+-}$, $2^{+-}$, and $1^{-+}$, are much higher in mass. It 
is therefore difficult to produce glueballs with exotic quantum numbers.
To disentangle 
glueballs with conventional quantum numbers from a dense background of 
conventional states is a painstaking task.  

\begin{figure}[ht]
\begin{center}
\includegraphics[width=8.0cm]{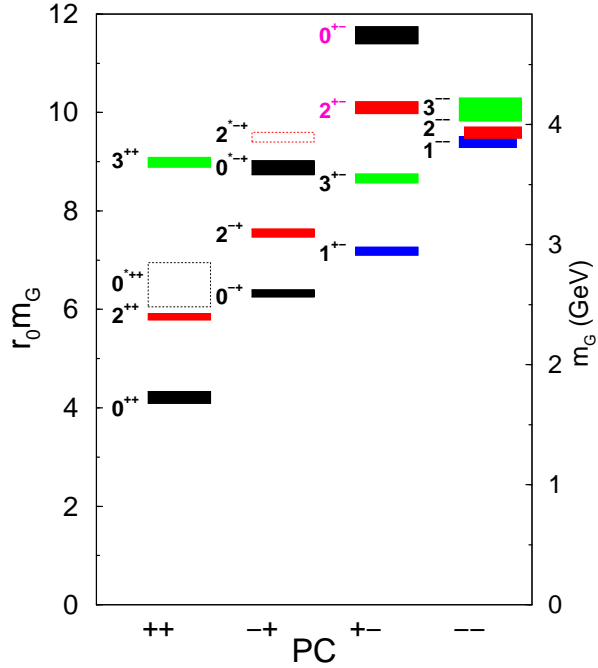} 
\caption{The mass of the glueball states.  The scale is set by 
$r_0$ with $1/r_0=410(20)$~MeV. From Morningstar and Peardon \cite{mp97}. }
\end{center}
\end{figure}

\subsection{Glueball Decays}

We expect glueball decays to have flavour symmetric couplings to final 
state hadrons:
\begin{equation}
{{\Gamma(G\to \pi\pi: \; K\bar{K}: \; \eta\eta : \; \eta\eta' : \; 
\eta'\eta')}\over{\hbox{Phase Space}}} \simeq 3:4:1:0:1
\label{eqn:gluedec}
\end{equation}
The situation is complicated by mixing with $q\bar{q}$ and 
$q\bar{q}q\bar{q}$ so the physical states are linear combinations:
\begin{equation}
\ket{f_0}=\alpha \ket{n\bar{n}} +\beta \ket{s\bar{s}} + \gamma \ket{G} 
+\delta \ket{q\bar{q}q\bar{q}}
\end{equation}
Mixing will both shift the unquenched glueball masses and distort the 
naive patterns of couplings given by eqn. (\ref{eqn:gluedec}) 
\cite{ca96,close00}.

Meson properties can be used to extract the mixings and understand the 
underlying dynamics.  For example, central production of the isoscalar 
scalar mesons has found the ratio of partial widths 
 $\Gamma(K\bar{K})/\Gamma(\pi\pi)$ to be \cite{wa102a}:
\begin{eqnarray}
f_0(1370) & <1 & (0.5 \pm 0.2)  \nonumber \\
f_0(1500) &  << 1 &  (0.3\pm 0.1) \\
f_0(1710) & >> 1 & (5.5 \pm 0.8)  \nonumber 
\end{eqnarray}
Relating this information to theoretical expectations Close and Kirk 
find \cite{close00}:
\begin{eqnarray}
\ket{f_0(1370)} & = & 
-0.79 \ket{n\bar{n}} - 0.13 \ket{s\bar{s}} + 0.60 \ket{G} \nonumber \\
\ket{f_0(1500)} & = & 
-0.62 \ket{n\bar{n}} + 0.37 \ket{s\bar{s}} - 0.69 \ket{G}  \\
\ket{f_0(1710)} & = & 
+0.14 \ket{n\bar{n}} +0.9 \ket{s\bar{s}} + 0.39 \ket{G} \nonumber
\end{eqnarray}
A similar analysis was done by Amsler \cite{amsler02}.  The point is 
not the details of a specific mixing calculation but that mixing is an 
important consideration that must be taken into account in the 
phenomenology.

Before proceeding to hadronic production of glueballs we mention that
two photon couplings are a sensitive probe of $q\bar{q}$ content 
\cite{close00}. 
The L3 collaboration at LEP sees the $f_0(1380)$ and $ f_0(1710)$ in 
$\gamma\gamma \to K\bar{K}$ but not the $f_0(1500)$. 
Because gluons do not carry electric charge, glueball 
production should be suppressed in $\gamma\gamma$ collisions.  
Quite some time ago Chanowitz \cite{chanowitz}
quantified this in a parameter he called 
``stickiness'' given by the ratio of meson production 
in radiative $J/\psi$ decay to two photon couplings:
\begin{equation}
S= {{\Gamma (J/\psi \to \gamma X)}\over{PS(J/\psi \to \gamma X)}}
\times {{PS(\gamma\gamma \to X)}\over {\Gamma(\gamma\gamma \to X)}}
\end{equation}
where $PS$ denotes phase space.  
A large value of $S$ is supposed to reflect an enhanced glue content.

\subsection{Glueball Production}

There are three processes which are touted as good places to look for 
glueballs:
\begin{enumerate}
\item $J/\psi \to \gamma X$
\item $p\bar{p}$ annihilation
\item $pp\to p_f(G)p_s$ central production
\end{enumerate}
It is the latter process that is relevant to COMPASS.
Central production is understood to proceed via gluonic pomeron 
exchange.  It is expected that glueball production has to compete with 
$q\bar{q}$ production. However,  a kinematic filter has been proposed 
which appears to suppress established $q\bar{q}$ states when in a 
P-wave or higher wave \cite{ck97}.

In the central production process:
\begin{equation}
pp\to p_f(G)p_s
\end{equation}
$p_s$ and $p_f$ represent the slowest and fastest outgoing protons.  
Central production is believed to be dominated by double {\it pomeron} 
exchange.  The pomeron is believed to have a large gluonic content.  
Folklore assumed that the pomeron has $J^{PC}=0^{++}$ quantum numbers 
and therefore gives rise to a flat distribution.  But the distribution 
turns out not to be flat and is well modelled assuming a $J=1$ 
exchange particle \cite{wa102}.  In other words the pomeron transforms as a 
non-conserved vector current.  Data from CERN experiment WA102 appears 
to support this hypothesis.

Close and Kirk \cite{ck97} have found a kinematic filter that seems to 
suppress established $q\bar{q}$ states when they are in $P$ and higher 
waves.  The pattern of resonances depends on the vector difference of 
the transverse momentum recoil of the final state protons:
\begin{equation}
dP_T= |\vec{k}_{T_1}-\vec{k}_{T_2}|
\end{equation}
For $dP_T$ large, the well established $q\bar{q}$ states are prominent 
while for $dP_T$ small, the established $q\bar{q}$ states are 
suppressed and the $f_0(1500)$, $f_0(1710)$, and $f_0(980)$ survive.

\begin{figure}[ht]
\begin{center}
\includegraphics[width=10.0cm]{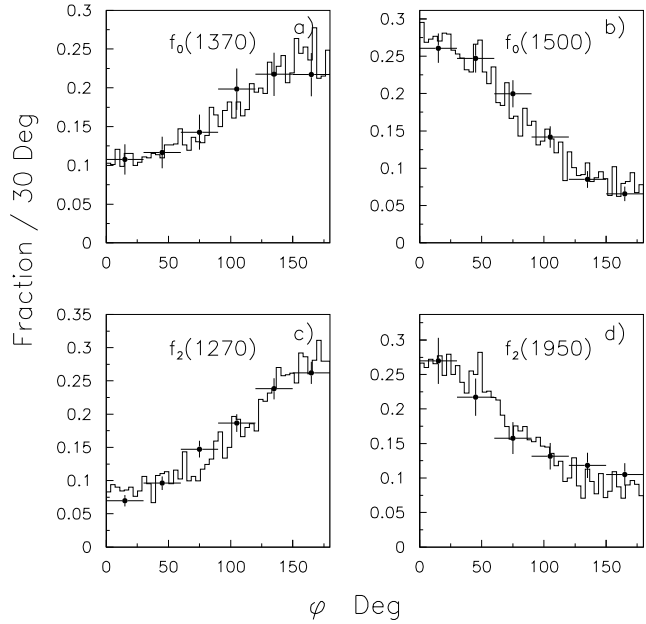} 
\caption{The $\phi$ distributions for (a) $f_0(1370)$ (b) $f_0(1500)$
(c) $f_2(1270)$ (d) $f_2(1950)$ for the data (dots) and the Monte 
Carlo (histogram).  From Close and Schuler \cite{cs99}.}
\end{center}
\end{figure}

Close, Kirk, and Schuler give a good account of the data by modeling the 
pomeron as a nonconserved vector exchange  \cite{cs99}.  They find that 
the $\phi$ angular distribution, the angle between the $k_T$ 
vectors, appears to distinguish between the production of different 
states \cite{wa102,cs99}.  In particular:
\begin{description}
\item[$0^{-+}$] Parity requires the vector pomeron to be transversely 
polarized.  The distribution peaks at $90^o$.
\item[$1^{++}$] One pomeron is transverse and the other longitudinal 
and the distribution peaks at $180^o$.
\item[$2^{-+}$] Similar to the $0^{-+}$ case but peaks at $0^o$. 
Helicity 2 is suppressed by Bose statistics.
\item[$2^{++}$] Established states peak at $180^o$ while the 
$f_2(1950)$ peaks at $0^o$.
\item[$0^{-+}$] Some states peak at $0^o$ while others are spread out:
\begin{itemize} 
\item $f_0(1500)$, $f_0(1710)$, and $f_0(980)$ peak at small $\phi$.
\item $f_0(1370)$ peaks at large $\phi$.
\end{itemize}
\end{description}
The fact that the $f_0(1370)$ and $f_0(1500)$ have different $\phi$ 
dependence indicates that it is not just a $J$ dependent phenomena
\cite{wa102b,wa102c}.

The $0^{++}$ and $2^{++}$ expect both $TT$ and $LL$ contributions.  
The differential cross section is given by \cite{cs99,cks00}:
\begin{equation}
{{d\sigma}\over {d\phi}} \sim \left[ { 1+ {{\sqrt{t_1 t_2}} 
\over{\mu^2}} {a_t\over a_L} \cos\phi } \right]^2
\end{equation}
Differential cross sections for scalar and tensor mesons
are shown in Fig. 2 \cite{cs99,cks00}.
Good fits to the distributions are obtained by varying 
$\mu^2 a_L/a_T$  with $\mu^2 a_L/a_T$ $=-0.5$~GeV$^2$ for $f_0(1370)$,
$=+0.7$~GeV$^2$ for $f_0(1500)$,
$=-0.4$~GeV$^2$ for $f_2(1270)$, and
$=+0.7$~GeV$^2$ for $f_0(1950)$.  Thus, the $\phi$ distributions are 
fitted with only 1 parameter.

\section{HYBRID MESONS}

Hybrid mesons are defined as those in which the gluonic component is 
non-trivial.  There are two types of hybrids; vibrational hybrids and 
topological hybrids.  The hybrid spectrum is generated by generating 
effective potentials from adiabatically varying gluonic flux tubes.  A 
given adiabatic surface corresponds to some string topology and 
excitation.  This is illustrated in Fig. 3.
In the flux-tube model the lowest excited adiabatic 
surface corresponds to transverse excitations of the flux tube. 

\begin{figure}[ht]
\begin{center}
\includegraphics[width=10.0cm]{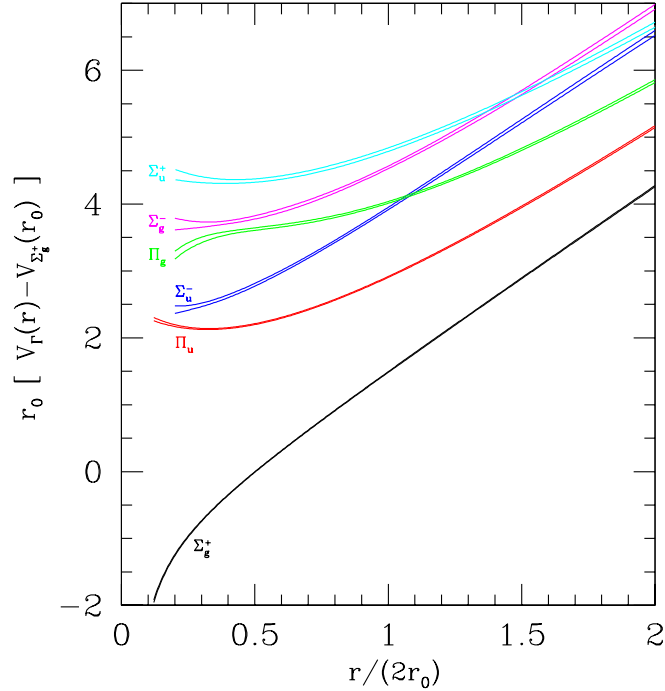} 
\caption{
A set of hybrid adiabatic surfaces for static central 
potentials.  $\Lambda = \Sigma ,\; \Pi , \; \Delta , \ldots$ 
corresponds to the magnitude of $J_{glue}=0, \; 1, \; 2, \ldots$ 
projected onto the molecular axis.  The superscript $=\pm$ corresponds 
to the even or oddness under reflections in a plane containing the 
molecular axis and the subscript $u/g$ corresponds to odd/even charge 
conjugation plus spatial inversion about the midpoint.  The familiar 
$q\bar{q}$ potential is labelled as $\Sigma^+_g$ and the first-excited 
potential is the $\Pi_u$ so the lowest lying hybrid mesons should 
be based on this potential. 
The double lines on the excited surfaces indicate the calculational 
uncertainty in determining the potential.  From Juge, Kuti and 
Morningstar \cite{juge98}.}
\end{center}
\end{figure}

While this picture is appropriate for heavy quarkonium it is not at all 
clear that it can be applied to light quark hybrids.  Nevertheless, 
given that the constituent quark model works so well for light quarks, 
it is not unreasonable to also extend the flux tube description to light 
quarks.
In the flux tube model the lowest mass hybrid mesons with light quark 
content have masses $\sim 1.9$~GeV \cite{ip85,ikp85,bcs95}.  
There is a double degeneracy with 
$J^{PC}=0^{+-}$, $0^{-+}$, $1^{+-}$, $1^{-+}$, $2^{+-}$, $2^{-+}$,
$1^{++}$, $1^{--}$ corresponding to the two transverse polarizations 
of the flux tube.  The degeneracies are expected to be broken by the 
different excitation energies of the flux tube modes, spin dependent 
effects, and mixings with conventional $q\bar{q}$ states (and 
possibly $q\bar{q}q\bar{q}$).  Lattice results are generally 
consistent with these predictions with $M(1^{-+})\sim 1.9$~GeV,
$M(0^{+-})\sim 2.1$~GeV, and $M(2^{+-})\sim 2.1$~GeV 
\cite{michael99,lat-hyb}.

\subsection{Hybrid Meson Decays}

Decay properties are a crucial tool in both directing  
exotic hybrid meson searches and 
to distinguish hybrids with conventional quantum 
numbers from conventional $q\bar{q}$ states.
A general selection rule 
for hybrid decays, which appears to be universal to all models,
is that to preserve the symmetries of quark and colour fields about 
the quarks, the $\Pi_u$ hybrid must decay to a P-wave meson 
\cite{page97,mmp02}.  
In other 
words it cannot transfer angular momentum to relative angular momentum 
between final state mesons but rather, to internal angular 
momentum of one of the final state mesons.  For the case of the 
$1^{-+}$ exotic the $\hat{\rho}\to b_1 \pi , \; f_1 \pi$ modes are 
expected to dominate.

To calculate hybrid properties we need to rely on models.  We will use 
the results of the flux tube model \cite{ikp85,cp95} which is based on strong 
coupling Hamiltonian lattice QCD.  The degrees of freedom 
are quarks and flux-tubes.  This model provides a unified framework 
for conventional hadrons, multiquark states, hybrids, and glueballs.

The flux-tube model predictions of Close and Page
for the dominant decay widths of exotic hybrid mesons are given in Table 1 
\cite{cp95}.  
One can see that the $\hat{a}_0$ and $\hat{f}'_0$ are too broad to be 
observed as resonances. 
The $\hat{\omega}_1$ decays to  $a_1 \pi$ and 
$K_1 K$ as does the $\hat{\phi}_1$.  These final states are 
notoriously difficult to reconstruct.  Thus the best bets for finding 
exotic hybrids are the decays $\hat{\rho}_1\to [b_1\pi]_S, \; 
[f_1\pi]_S$ with $\Gamma \simeq 170$~MeV.
This is why the $\hat{\rho}$ is the focus of 
so much attention in hybrid searches.  
The narrow $f_1(1285)$ provides a particularly useful tag in 
$\hat{\rho}\to \pi f_1$.
Although there is a general 
consensus among models with respect to the qualitative properties 
given here one should be aware that there is some disagreement in  
predictions.  See, for example, the predictions of Page, Swanson and 
Szczepaniak \cite{ss,pss}.  In particular, Page {\it et al.} \cite{pss}
predict 
the $\hat{a}_2$ width to be very narrow so that it would be useful to 
search for $a_2 \pi$ and $h_1 \pi$ final states.  If nothing else this 
would be a good test of the models.

\begin{table}[t]
\begin{center}
Table 1: Dominant decay widths of exotic hybrid mesons. From 
Close and Page \cite{cp95}.
\vskip0.2cm
\begin{tabular}{|l|l|c|r|}
\hline
Initial State & Final State & L & $\Gamma$ \\
\hline
$\hat{\rho} (1^{-+})$ 	& $b_1(1235) \pi$ 	& S & 100 \\
		& 			& D & 30 \\
		& $f_1(1285) \pi$ 	& S & 30 \\
		& 			& D & 20 \\
\hline
$\hat{\omega} (1^{-+})$ 	& $a_1(1260) \pi$ 	& S & 100 \\
			& 			& D & 70 \\
		& $K_1(1400) K$ 	& S & 100 \\
\hline
$\hat{\phi} (1^{-+})$ & $K_1(1270) K$ 	& D & 80 \\
		& $K_1(1400) K$ 	& S & 250 \\
\hline
$\hat{a}_2 (2^{+-})$ 	& $a_2(1320) \pi$ 	& P & 450 \\
		& $a_1(1260) \pi$ 	& P & 100 \\
		& $h_1(1170) \pi$ 	& P & 150 \\
\hline
$\hat{f}_2 (2^{+-})$ 	& $b_1(1235) \pi$ 	& P & 500 \\
\hline
$\hat{f}_2' (2^{+-})$ & $K_2^*(1430) K$ 	& P & 250 \\
		& $K_1(1400) K$ 	& P & 200 \\
\hline
$\hat{a}_0 (0^{+-})$ 	& $a_1(1260) \pi$ 	& P & 800 \\
		& $h_1(1170) \pi$ 	& P & 100 \\
\hline
$\hat{f}_0 (0^{+-})$ 	& $b_1(1235) \pi$ 	& P & 250 \\
\hline
$\hat{f}_0' (0^{+-})$ & $K_1(1270) K$ 	& P & 800 \\
		& $K_1(1400) K$ 	& P & 50 \\
\hline
\end{tabular}
\end{center}
\end{table}

Although hybrid mesons with exotic quantum numbers give a distinctive 
signature, hybrids with conventional quantum numbers are also expected 
in the meson spectrum.  The situation is more complicated than simply 
looking for additional states because we expect strong mixing 
between non-spin exotic hybrids and conventional mesons with the same 
quantum numbers.  Thus, to distinguish non-exotic hybrids from 
conventional states requires detailed predictions of properties 
\cite{bcps,cp95,cp97,dk99}.  

A first example is whether the  $\pi(1800)$ is a conventional $3S$ 
isovector pseudoscalar meson (the 2nd radial excitation of the $\pi$) 
or a hybrid meson.  Predictions for the partial width of a $\pi_{3S}$ 
and $\pi_H$ are given in Table 2. The flux tube model predicts that 
the $\pi_{3S}$ decays to $\omega\pi$ but the $\pi_H$ does not.  
Likewise, the $\pi_H$ has a large partial width to $f_0(1300)\pi$ 
while for the $\pi_{3S}$ this partial width is quite small.  
Therefore the 
$\rho\omega$ and $f_0(1300)\pi$ modes can be used as discrimators
between the two possibilities.
The $\pi_{3S}$ has been observed in $\pi f_0(1300)$ lending support to its 
identification as a hybrid. 
\begin{table}[h]
\begin{center}
Table 2: Partial decay widths for the $\pi(3S)$ and $\pi_H$.
From Barnes {it et al.} \cite{bcps}.
\vskip0.2cm
\begin{tabular}{|l|c|c|c|c|c|c|c|}
\hline
State 	& \multicolumn{6}{|c|}{Partial widths to final states} & Total \\
\cline{2-7}
	& $\pi\rho$ & $\omega\rho$ & $\rho(1465)\pi$ & $f_0(1300)\pi$ 
		& $f_2 \pi$ & $K^* K$ & \\
\hline
$\pi_{3S}(1800)$ & 30 & 74 & 56 & 6 & 29 & 36 & 231 \\
$\pi_H(1800)$ & 30 & --- & 30 & 170 & 6 & 5 & $\sim 240$ \\
\hline
\end{tabular}
\end{center}
\end{table}

Another example is that of the $\rho'$ and $\omega'$ mesons.  One 
expects the physical vector mesons to be a linear combination 
\begin{equation}
\ket{V} = \sum_n \alpha_n \ket{n^3S_1} + \sum_m \beta_m \ket{m^3D_1} + 
\gamma \ket{V_H}
\end{equation}
To disentangle the various components of the physical mesons we need 
to perform a detailed comparison between the observed states and the 
predictions for the unmixed $q\bar{q}$ and $V_h$ states, much as was 
done for the scalar iso-scalar mesons.  Partial width 
predictions are shown in Table 3
for the $\rho_{2S}(1465)$, $\rho_{1D}(1700)$, and
$\rho_H(1500)$ states.  For this example the $\pi h_1$ and $\pi a_1$ 
decay modes can discriminate between the $\rho_{2S}$, $\rho_{1D}$ and 
$\rho_H$ to disentangle the mixings.

\begin{table}[h]
\begin{center}
Table 3: Partial decay widths for the $\rho_{2S}$, $\rho_{1D}$ and 
$\rho_H$. From Barnes {it et al.} \cite{bcps}.
\vskip0.2cm
\begin{tabular}{|l|c|c|c|c|c|c|c|c|c|}
\hline
State 	& \multicolumn{8}{|c|}{Partial widths to final states} & Total \\
\cline{2-9}
	& $\pi\pi$ & $\omega\pi$ & $\rho\eta$ & $\rho\rho$ 
		& $KK$ & $K^* K$ & $h_1\pi$ & $a_1\pi$ & \\
\hline
$\rho_{2S}(1465)$ & 74 & 122 & 25 & -- & 35 & 19 & 1 & 3 & 279 \\
$\rho_{1D}(1700)$ & 48 & 35 & 16 & 14 & 36 & 26 & 124 & 134 & 435 \\
$\rho_H(1500)$ & 0 & 5 & 1 & 0 & 0 & 0 & 0 & 140 & $\sim 150$ \\
\hline
\end{tabular}
\end{center}
\end{table}

A similar excercise can be applied to the isocalar sector with the 
relevant partial widths given in Table 4. The decays $\omega(1420)\to 
\pi b_1$ and $\omega(1600) \to \pi b_1$ are both observed to be small 
so neither is likely to be a pure $1^3D_1$ state.  This implies that 
one is the $2^3S_1$ and indicates that the other has significant
$\omega_H$ content.  It is clearly important to find the 3rd state in 
this set and determine some of the other branching ratios.  
The essential point is that although the two states may have the same 
$J^{PC}$ quantum numbers they have different internal structure which 
will manifest itself in their decays.  
Unfortunately,  nothing is simple and we once again point out that 
strong mixing is expected between hybrids with conventional quantum 
numbers and $q\bar{q}$ states with the same $J^{PC}$ so that the decay 
patterns of physical states may not closely resemble those of either 
pure hybrids or pure $q\bar{q}$ states. 
With 
enough information one could perform an analysis similar to the one 
performed on the scalar meson sector by Close and Kirk \cite{close00}.

\begin{table}[t]
\begin{center}
Table 4: Partial decay widths for the $\omega_{2S}$, $\omega_{1D}$ and 
$\omega_H$. From Barnes {it et al.} \cite{bcps}.
\vskip0.2cm
\begin{tabular}{|l|c|c|c|c|c|c|}
\hline
State 	& \multicolumn{5}{|c|}{Partial widths to final states} & Total \\
\cline{2-6}
	& $\rho\pi$ & $\omega\eta$ & $KK$ & $K^* K$ & $b_1\pi$ &  \\
\hline
$\omega_{2S}(1419)$ & 328 & 12 & 31 & 5 & 1 & 378 \\
$\omega_{1D}(1649)$ & 101 & 13 & 35 & 21 & 371 & 542 \\
$\omega_H(1500)$ & 20 & 1 & 0 & 0 & 0  & $\sim 20$ \\
\hline
\end{tabular}
\end{center}
\end{table}


\subsection{Production of Hybrid Mesons}

Hybrid mesons can be produced in a number of processes:
\begin{enumerate}
\item $J/\psi \to \gamma X$
\item $\bar{p}p$ annihilation
\item peripheral production
\item photoproduction
\end{enumerate}
It is the latter two processes which are relevant to the COMPASS 
collaboration.  Peripheral production is discussed in more detail by 
Dorofeev \cite{dorofeev} and photoproduction by Moinester 
\cite{moinester} in these proceedings.

\subsubsection{Hadronic Peripheral Production}

In peripheral production the beam particle is excited and
exchanges momentum and quantum numbers with the target nucleus via an 
exchange particle.  The excited meson continues to move forward,
subsequently decaying into the decay products which are detected by the 
experiment.  This is shown schematically in Fig. 4.
Examples of experiments which studied peripheral production 
are LASS at SLAC, E852 at Brookhaven, 
BENKEI at KEK, VES at 
IHEP/Serpukhov 
and GAMS at  CERN.

\begin{figure}[ht]
\begin{center}
\includegraphics[width=6.0cm]{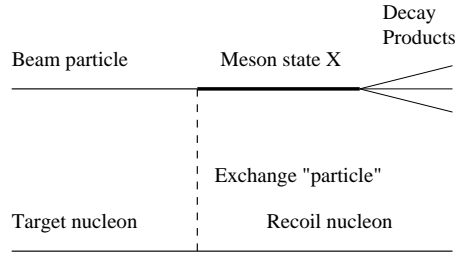} 
\caption{Peripheral production of mesons. }
\end{center}
\end{figure}

Evidence for hybrid mesons has been seen by the VES collaboration 
\cite{ves93} in $\rho^0\pi^-$, $\pi\eta$, and $\pi b_1$ final states 
in the reaction 
\begin{equation}
\pi^- N \to (\eta\pi^+\pi^-)\pi^- N \hbox{ with a 37 GeV/c } \pi 
\hbox{ beam}
\end{equation}
and by BNL E852 \cite{adams98} in the final state $\pi f_1(1285)$  in 
the reaction
\begin{equation}
\pi^- p \to (\pi^-\pi^+\pi^-)p \hbox{ with a 18 GeV/c } \pi \hbox{ beam.}
\end{equation}

There is no reason a priori to expect that any type of hadron is 
preferred over any other in this mechanism.  The $\pi$ exchange 
mechanism only provides access to natural parity states.  But the 
advantage of very high statistics is that
with enough statistics one could use $t$-distributions to distinguish 
between different exchange particles which would allow one to study 
states other than the natural parity states.

E852 at Brookhaven provides a nice lesson of the advantages of high 
statistics \cite{adams98}.  
In Fig. 5 the event rates for $\pi^- p \to \pi^+\pi^-\pi^- 
p$ at 18 GeV/c is shown as a function of 
$\pi^+\pi^-\pi^-$ invariant mass.  Structure is seen corresponding to 
the $a_1$, $a_2$, and $\pi_2$ mesons although it would be difficult to 
draw conclusions from this figure alone.  However, with the large data 
sample a partial wave analysis can be performed.  The results are 
also shown in Fig. 5.  One now sees clear resonances corresponding to the 
$a_1$, $\pi_2$ and $a_2$.  These reference waves can be used to 
measure the phase shift of the exotic waves that are being looked 
for.  This is shown in Fig. 6 where intensity and phase of the 
$1^{-+}$ exotic signal clearly stands out.

The lesson is that a PWA is a necessary component of any study of 
meson physics and that high statistics offer the opportunity to 
perform the necessary studies.

\begin{figure}[t]
\centerline{
\begin{minipage}[t]{6.0cm}
\vspace*{13pt}
\centerline{\hspace{-1.0cm}
\includegraphics[bb=41 25 230 212,width=6.0cm,clip=true]{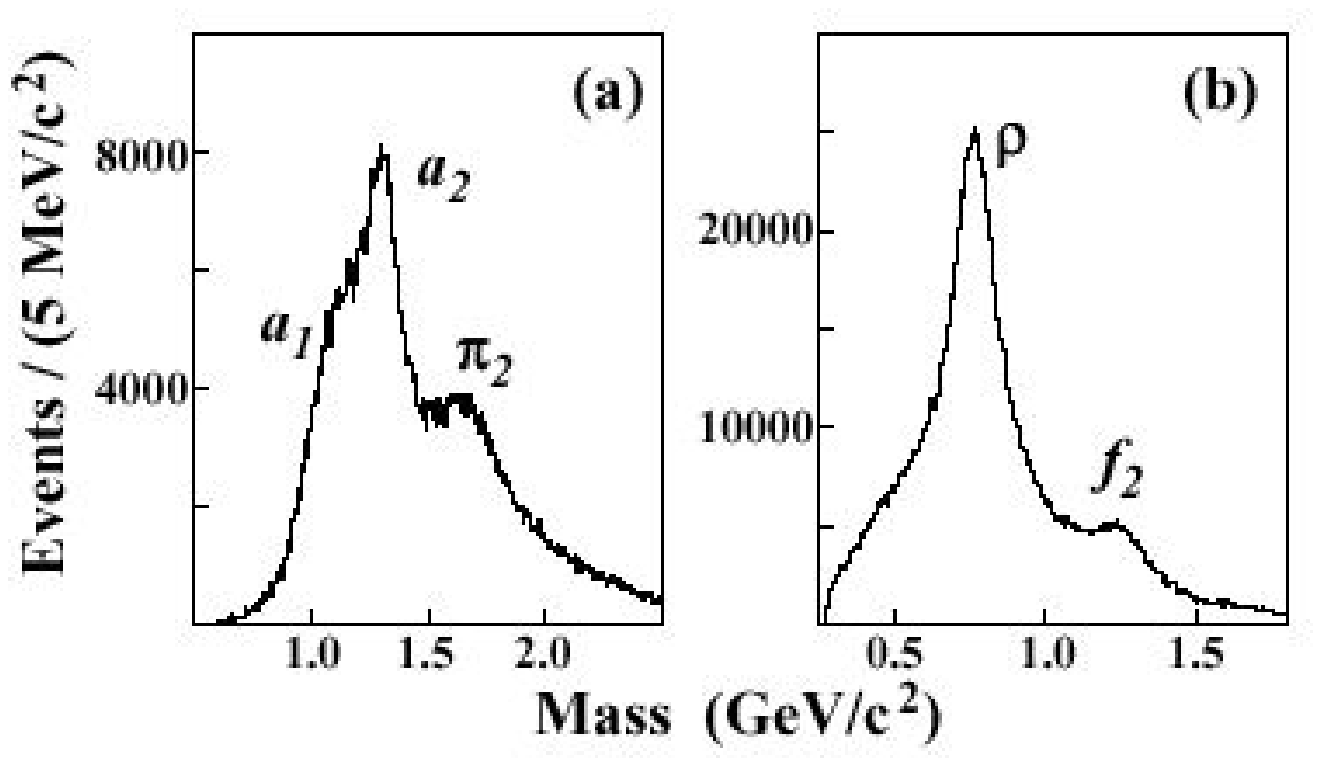} }
\vspace*{13pt}
\end{minipage} 
\hspace*{0.8cm}
\begin{minipage}[t]{8.0cm}
\vspace*{13pt}
\centerline{\hspace{-0.2cm}
\includegraphics[bb=0 0 473 390,width=8.0cm,clip=true]{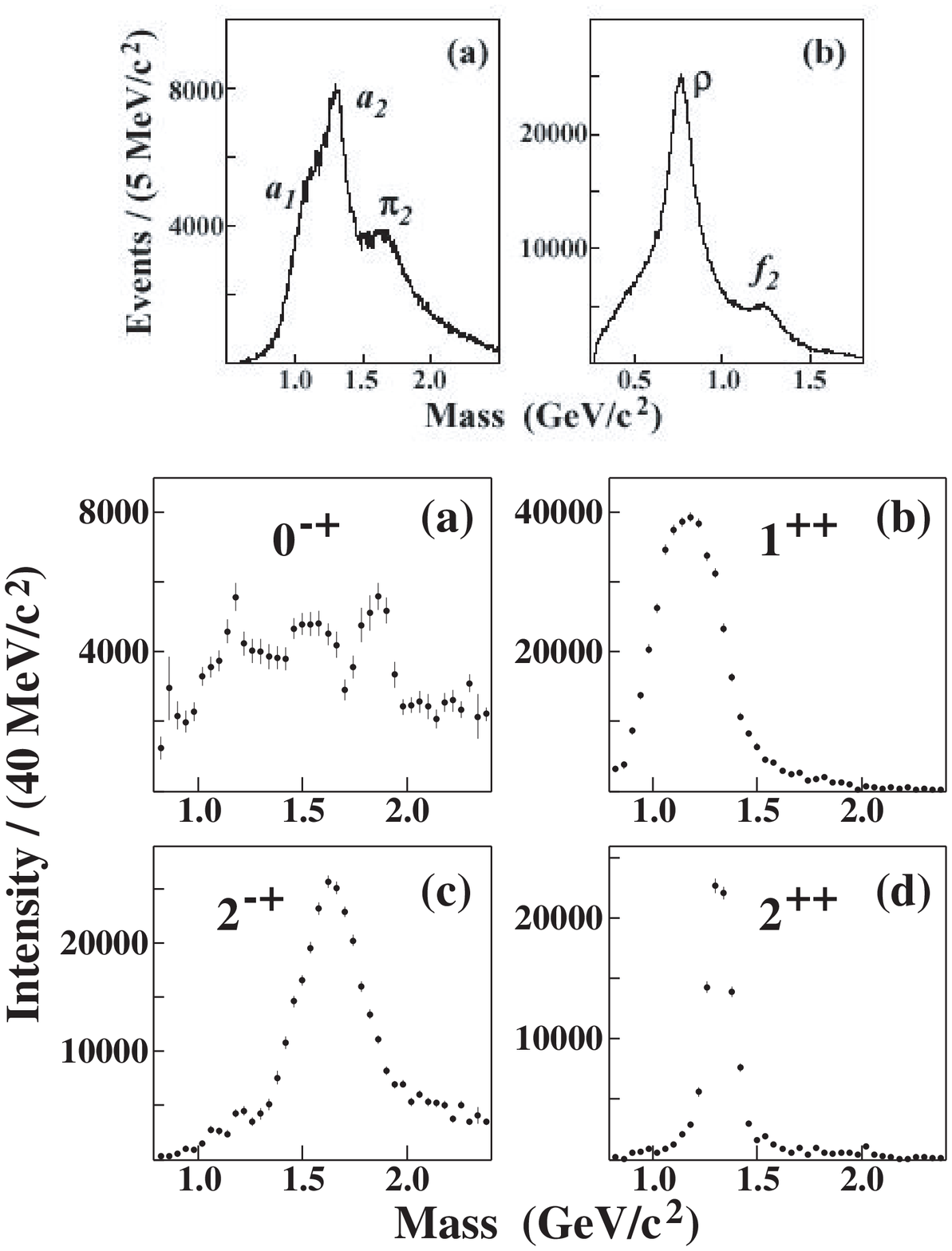} }
\vspace*{13pt}
\end{minipage}
}
\caption{Invariant mass distribution of $\pi^+\pi^-\pi^-$. The figure 
on the left shows the raw data and the figure on the right shows the 
results of a PWA. From Ref. \cite{adams98}. }
\end{figure}

\begin{figure}[h]
\begin{center}
\includegraphics[bb=0 248 242 432,width=6.0cm,clip=true]{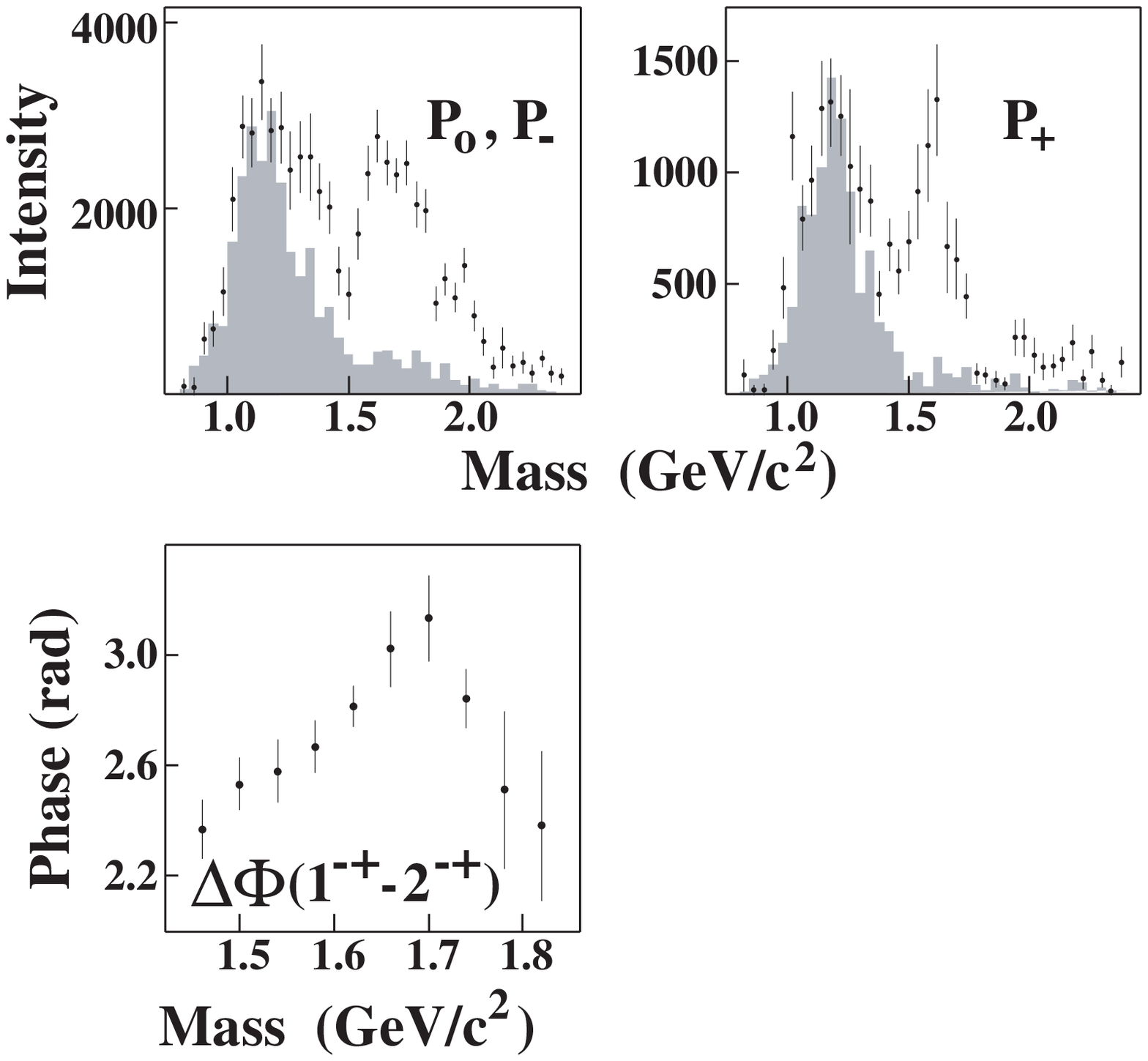} 
$\qquad$
\includegraphics[bb=0 34 241 215,width=6.0cm,clip=true]{exotic.eps} 
\caption{The figure on the left shows
the wave intensities of the $1^{-+}$ exotic waves.  The PWA fit 
to the data is shown as the points with error bars and the shaded 
histograms show estimated contributions from all nonexotic waves due 
to leakage.   The figure on the right shows the phase difference 
between the $1^{-+}$ and $2^{-+}$ waves from a
coupled mass-dependent Breit-Wigner fit.
From Ref. \cite{adams98}. }
\end{center}
\end{figure}

\subsubsection{Photoproduction}

COMPASS offers a unique opportunity in that it can also study hybrid 
meson production via photoproduction by way of initial muon beams.  
Photoproduction is qualitatively different to hadronic peripheral 
production so that the series of preferred excitations is likely to be 
different.  Additionally, it is a strong source of $s\bar{s}$ states.  
Via vector meson dominance one can view the photon as a linear 
combination of the $\rho$, $\omega$, $\phi$ and other vector mesons.
In vector mesons the quark spins are aligned in a $S=1$ triplet 
state.  As hybrid mesons with exotic quantum numbers are also in a 
spin triplet state it is believed that exotic hybrid mesons are 
favoured by this process.  At the present time there is virtually no 
photoproduction data available.  
Some time ago the Omega Photon Collaboration studied the process $\gamma 
p\to (b_1 \pi)p$ at 25-50~GeV incident energy witht the specific 
intention of seeking hybrids \cite{omega}.
The most recent photoproduction 
experiment was done at SLAC studying $\gamma p \to \pi^+\pi^+\pi^- n$ at 
19~GeV \cite{condo93}.  It showed hints of exotics but unfortunately, 
the statistics were rather low.
A dedicated high statistics experiments with the power of modern 
detection and analysis should reexamine this process \cite{cp95b}. 
This is almost virgin territory and an area to which the COMPASS collaboration 
could make important contributions.

\section{MULTIQUARK MESONS}

In addition to conventional $q\bar{q}$ mesons, hybrids and glueballs, 
multiquark mesons are also expected to exist. It was noted 
in the discusions of glueballs and hybrids 
that they contribute to the physical spectrum.

While there is no room to discuss this topic in any detail I mention it 
as an additional ingredient that one should be aware of when studying 
meson spectroscopy.  Several  examples exist of multiquark 
candidates.  It has long been believed that the $f_0(980)$ and 
$a_0(980)$ are multiquark states although their exact nature; a 
compact $q\bar{q}q\bar{q}$ object or an extended $K\bar{K}$ molecule 
is the focus of vigorous debate.  The nature of the $f_1(1430)$ is a 
longstanding puzzle and is part of our lack of understanding of what 
is known as the $E/\iota$ puzzle.  There is speculation that it is a 
$K^*K$ bound state.  

Multiquark states can also have exotic quantum numbers.  The best bets 
along this line of study would be fractional or doubly charged mesons 
although it has been speculated that at least one of the 
$J^{PC}=1^{-+}$ exotic candidates is a $\bar{q}q\bar{q}q$ object.

\section{SUMMARY}

The existence of non-$q\bar{q}$ mesons is the most important 
qualitative open question in QCD.  The discovery and mapping out of 
the glueball and hybrid meson spectrum is a crucial test of QCD.  It 
will help validate lattice QCD as an important computational tool for 
non-perturbative field theory.  It will take 
detailed studies to distinguish glueball and hybrid candidates from 
conventional $q\bar{q}$ states.  This will require extremely high 
statistics experiments to measure meson properties such as partial 
widths and  production mechanisms.  COMPASS is unique.  It has 
numerous tools to do this via $\pi$, $K$, $p$, and $\mu$ beams.  
COMPASS can make important advances in this field.  I strongly 
encourange you to do so.

\noindent

\section*{ACKNOWLEDGEMENTS}

The author thanks Frank Close for helpful comments in 
preparing this manuscript, the organizers of the 
workshop for providing a stimulating environment for the discussion of 
these issues, and the DESY theory group for their warm hospitality 
where this was written up.
This work was partially funded by the Natural Sciences and Engineering 
Research Council of Canada.

\end{document}